\documentclass{osa-article}

\usepackage{braket}

\usepackage{color}

\journal{oe}

\begin{document}

\title{Feedforward-enhanced Fock state conversion with linear optics}

\author{Vojt\v{e}ch \v{S}varc\authormark{1*}, Josef Hlou\v{s}ek\authormark{1}, Martina Nov\'{a}kov\'{a}\authormark{1}, Jarom\'{i}r Fiur{\'a}{\v{s}}ek\authormark{1}, and Miroslav Je\v{z}ek\authormark{1}}

\address{\authormark{1}Department of Optics, Faculty of Science, Palack\'y University, 17.\ listopadu 12, 77146 Olomouc, Czech Republic}
          
\email{\authormark{*}svarc@optics.upol.cz} 



\begin{abstract}
Engineering quantum states of light represents a crucial task in the vast majority of photonic quantum technology applications. Direct manipulation of the number of photons in the light signal, 
such as single-photon subtraction and addition, proved to be an efficient strategy for the task. Here we propose an adaptive multi-photon subtraction scheme where a particular subtraction task
 is conditioned by all previous subtraction events in order to maximize the probability of successful subtraction. We theoretically illustrate this technique on the model example of conversion of Fock states via photon subtraction. We also experimentally demonstrate the core building block of the proposal by implementing a feedforward-assisted conversion of two-photon state to a single-photon state.  
Our experiment combines two elementary photon subtraction blocks where the splitting ratio of the second subtraction beam splitter is affected by the measurement result from the first subtraction block 
  in real time using an ultra-fast feedforward loop. 
  The reported optimized photon subtraction scheme applies to a broad range of photonic states, including highly nonclassical Fock states and squeezed light, advancing the photonic quantum toolbox. 
\end{abstract}

\section{Introduction}

Preparation and controlled manipulation of nonclassical states of light lies at the heart of quantum optics and represents a key tool for the rapidly developing optical quantum technologies. 
Since the class of experimentally available deterministic unitary operations on quantum states of light is rather limited, it is extremely useful and fruitful to consider also probabilistic conditional 
operations that significantly extend the scope of quantum states that could be prepared, and transformations that could be implemented. The prime examples of such operations are the conditional single-photon 
addition and subtraction \cite{Wenger2004,Zavatta2004,Ourjoumtsev2006,
Nielsen2006,Wakui2007,Namekata2010,Bellini2010,Gerrits2010}. These elementary operations can be utilized to generate 
highly non-classical states with negative Wigner function \cite{Ourjoumtsev2006,Nielsen2006,Ourjoumtsev2009,
Nielsen10,Gerrits2010,Magana2019,Smith2013}, implement 
various optical quantum gates and operations \cite{Fiurasek2009,Marek2010,Tipsmark2011,Blandino2012,Coelho2016,Costanzo2017}, realize probabilistic noiseless quantum amplifiers 
\cite{Zavatta2011,Usuga2010}, distill continuous-variable entanglement \cite{Ourjoumtsev2007,Takahashi2010,Kurochkin2014} or to probe the fundamental properties of quantum mechanics 
\cite{Parigi2007,Zavatta2009}. Also many alternative schemes for optical quantum state preparation and manipulation via 
conditional single-photon or homodyne detection have been proposed and demonstrated \cite{Babichev2004,Bimbard2010,Huang2015,Yukawa2013,Sychev2017}.

Conditional photon subtraction can be performed by sending the light beam at a beam splitter that taps off a part of the signal that is subsequently measured with 
a single photon detector whose click heralds the photon subtraction.
In the experiments one usually employs a highly unbalanced beam splitter to reduce the negative influence of imperfect photon detection with non-unit efficiency $\eta$.
This makes the experimental photon subtraction closer to the action of annihilation operator $\hat{a}$, however at the expense of reduced success probability. 
With the advent of superconducting single-photon detectors and rapidly improved quantum detection efficiencies exceeding $90\%$ \cite{Nam2013,Laurat2016} it nevertheless becomes relevant to investigate also a different regime 
of approximate photon subtraction where one attempts to maximize the success probability by a suitable choice of the transmittances of the tapping beam splitters. 

It was shown previously that the success probability of 
photon subtraction from a travelling beam of light can be increased by a feedforward-controlled loop-based scheme \cite{Calsamiglia2001,Marek2018} where the 
light beam is repeatedly injected into the photon subtraction device until a subtraction event is detected. 
This approach enables asymptotically deterministic photon subtraction. The resulting subtraction operation depends on the input state 
and can be generally expressed as a mixture of quantum filters $t^{k\hat{n}}\hat{a}$ \cite{Marek2018}, where $\hat{n}$ is the photon number operator, $k$ is the number of loops the light beam has passed until a photon was subtracted, and 
$|t|<1$ is the beam splitter amplitude transmittance. In this paper we further investigate the advantages of feedforward-based photon subtraction and we consider general multi-photon subtraction schemes 
involving a sequence of several elementary photon subtraction blocks, where the transmittance of each tapping beam splitter is controlled by measurement results from all preceding elementary photon subtraction blocks.

Specifically, we study the model problem of conversion of a Fock state $|m\rangle$ to a Fock state $|n\rangle$ with $n<m$ by subtraction of $m-n$ photons. 
We consider a scheme involving $k$ elementary photon subtraction blocks and we demonstrate that the success probability of the scheme is maximized 
if we actively and adaptively choose a suitable transmittance of the beam splitter in $j$th subtraction block depending 
on the measurement outcomes of all the previous blocks. We find that this advantage of feedforward persists even for inefficient detectors and certain amount of optical losses.
We experimentally demonstrate this feedforward-based protocol for the conversion of a two-photon Fock state $|2\rangle$ to the single-photon state $|1\rangle$ 
using an electronically controlled variable fiber beam splitter formed by a Mach-Zehnder interferometer with electrooptics modulators placed in its arms \cite{Svarc2019}. 
Our measurement results clearly confirm the potential advantage of the feedforward-based photon conversion scheme. 

We note that efficient extraction of a single or several photons from a light beam can be also  implemented with the use of quantum light-atoms interactions \cite{Honer2011,Rosenblum2015}. 
The interaction of electromagnetic field in a cavity with atoms flying through it may also serve for quantum non-demolition photon counting with applications including observation of progressive quantum state collapse, 
preparation and stabilization of Fock states of the field and tomographic characterization of the cavity field states \cite{Gleyzes2007,Guerlin2007,Deleglise2008,Sayrin2011}. 
The atom based schemes are very promising but also very technologically demanding. 
Here we instead focus on simple and practicable all-optical setups with the goal to design schemes exhibiting high success probability while requiring only a few tunable beam splitters and single photon detectors. 

The rest of the paper is organized as follows. Theoretical description and analysis of the protocol is provided in Section 2. The experimental setup is described in Section 3 where also the experimental 
results are presented and discussed. Finally, Section 4 contains brief conclusions.

\section{Theory}
Here we present theoretical derivation of optimal feedforward-based schemes for conversion of optical Fock state $|m\rangle$ to Fock state $|n\rangle$ via subtraction of $m-n$ photons.
We first describe the method on the illustrative example of conversion of a two-photon Fock state $|2\rangle$ to the single-photon state $|1\rangle$ 
and then generalize the procedure to arbitrary $|m\rangle\rightarrow |n\rangle$ conversion with $m>n$.
The $|2\rangle\rightarrow |1\rangle$ conversion can be accomplished by single photon subtraction, whose simplest instance is depicted in Fig. 1(a). The input Fock state $|2\rangle$ 
impinges on a beam splitter with amplitude transmittance $t$ and reflectance $r$, where it is transformed into an entangled state of output spatial modes A and B,

\begin{equation}
|2\rangle_A|0\rangle_B \rightarrow t^2|2\rangle_A|0\rangle_B+\sqrt{2}tr|1\rangle_A|1\rangle_B+r^2|0\rangle_A|2\rangle_B.
\end{equation}
The output auxiliary mode B is measured with a photon number resolving detector. The conversion is successful if a single photon is detected, which occurs with probability $P(2,1)=2T(1-T)$, where $T=t^2=1-r^2$.
This probability is maximized for $T=\frac{1}{2}$ and we get $P_{\mathrm{max}}(2,1)=\frac{1}{2}$. If the detector on output mode B detects two photons, then the input state is destroyed and cannot be recovered. 
However, if no photons are detected, then we know that the output state of mode A is still the two-photon Fock state $|2\rangle$ and we can attempt to repeat the photon subtraction. 
The resulting feedforward-enhanced scheme is shown in Fig. 1(b). A second beam splitter and detector are placed after the first beam splitter, 
and the transmittance of the second beam splitter is controlled by the feedforward signal from the first detector. Let $T_1$ and $T_2$ 
denote the intensity transmittances of the first and second beam splitter, respectively. If the first detector detects one photon, then $T_2$ is set to $1$. 
On the other hand, if the first detector detects no photons, then $T_2$ is set to $\frac{1}{2}$. The overall success probability of this two-stage conversion scheme reads
\begin{equation}\label{eq:21conversion}
P(2,1|2)=T_1^2\times \frac{1}{2}+2T_1(1-T_1)\times 1= 2T_1-\frac{3}{2}T_1^2.
\end{equation}
This probability is maximized for $T_1=\frac{2}{3}$, and we get $P_{\mathrm{max}}(2,1|2)=\frac{2}{3}$.

\begin{figure}[t!]
\centering\includegraphics[width=\linewidth]{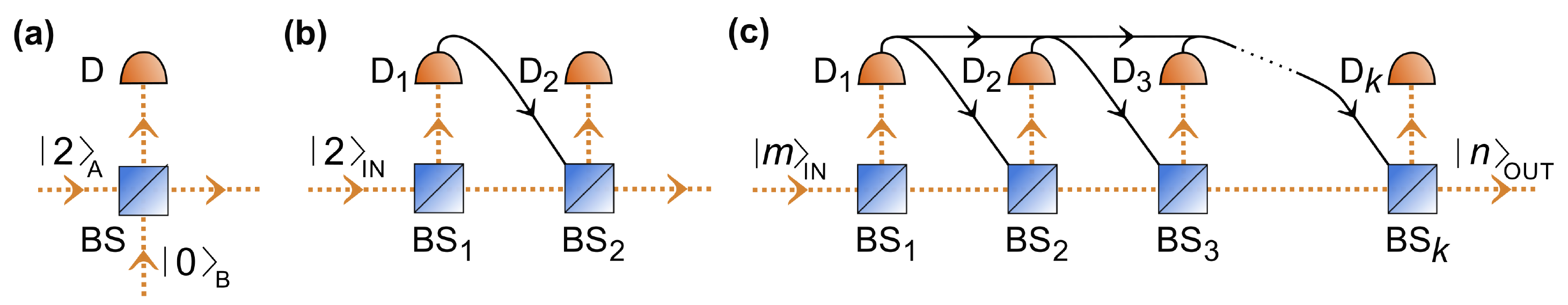}
\caption{Fock state conversion schemes. (a) The simplest way of $|2\rangle\rightarrow |1\rangle$ conversion utilizing single beam splitter. 
(b) Feedforward-enhanced $|2\rangle\rightarrow |1\rangle$ conversion using two beam splitters. (c) Generic scheme for $|m\rangle\rightarrow |n\rangle$ conversion exploiting $k$ beam splitters.}
\label{fig:schema}
\end{figure}

We now present a generic protocol for $k$-step conversion of Fock state $|m\rangle$ to Fock state $|n\rangle$, where $m>n$ and $k\geq 1$ denotes the number of elementary photon subtraction steps. The scheme is illustrated 
in Fig. 1(c) where the transmittance $T_j$ of beam splitter $\mathrm{BS}_j$ is controlled by the measurement outcomes of all preceding detectors $\mathrm{D}_l$, $l<j$. 
Let $P_{\mathrm{max}}(m,n|k)$ denote the maximum achievable conversion probability with $k$ steps. Suppose that we have found all optimal probabilities $P_{\mathrm{max}}(m-j,n|k-1)$. 
Considering the scheme in Fig. 1(c) as a combination of the first beam splitter and detector and a block performing optimal feedforward controlled 
conversion with $k-1$ steps, we can write
\begin{equation}
P(m,n|k)=\sum_{j=0}^{m-n} {{m}\choose{j}} T_1^{m-j}(1-T_1)^{j} P_{\mathrm{max}}(m-j,n|k-1).
\end{equation}
The optimal transmittance $T_1$ can be determined by finding roots of the polynomial
\begin{equation}
\frac{d P(m,n|k)}{d T_1}=0
\end{equation}
and choosing the root that lies in the interval $[0,1]$ and maximizes $P(m,n|k)$. Explicitly, the polynomial equation reads
\begin{equation}
\label{eqpolynomial}
\sum_{j=0}^{m-n} {{m}\choose{j}} T_1^{m-j-1}(1-T_1)^{j-1}[(m-j)(1-T_1)-jT_1] P_{\mathrm{max}}(m-j,n|k-1)=0.
\end{equation}
There are $n-1$ non-optimal roots $T_1=0$ and if we divide the equation (\ref{eqpolynomial}) by $T_1^{n-1}$ we end up with a polynomial equation of order $m-n$ 
that can be solved either numerically or even analytically for $m-n \leq 4$.
 
The optimal probability $P_{\mathrm{max}}(m,n|k)$ can be calculated iteratively. It is useful to formally define $P_{\mathrm{max}}(m,n|0)=0$, $m>n>0$, and we also have that $P_{\mathrm{max}}(m,m|k)=1$. 
Specifically, we first determine $P_{\mathrm{max}}(n+1,n|j)$ starting from $j=1$ and proceeding up to $j=k$. We then continue with determination of $P_{\mathrm{max}}(n+2,n|j)$, $1\leq j \leq k,$ 
and we repeat the whole calculation for all $P_{\mathrm{max}}(n+l,n|j)$ with increasing $l$ until we reach $n+l=m$. In Fig. 2 we plot the optimal probabilities $P_{\mathrm{max}}(m,n|k)$ 
for $6$ combinations of $m$ and $n$ and for up to $9$ subtraction steps. We can see that the conversion probability increases with the number of steps $k$ and asymptotically approaches $1$. 

\begin{figure}[t]
\centerline{\includegraphics[width=0.95\linewidth]{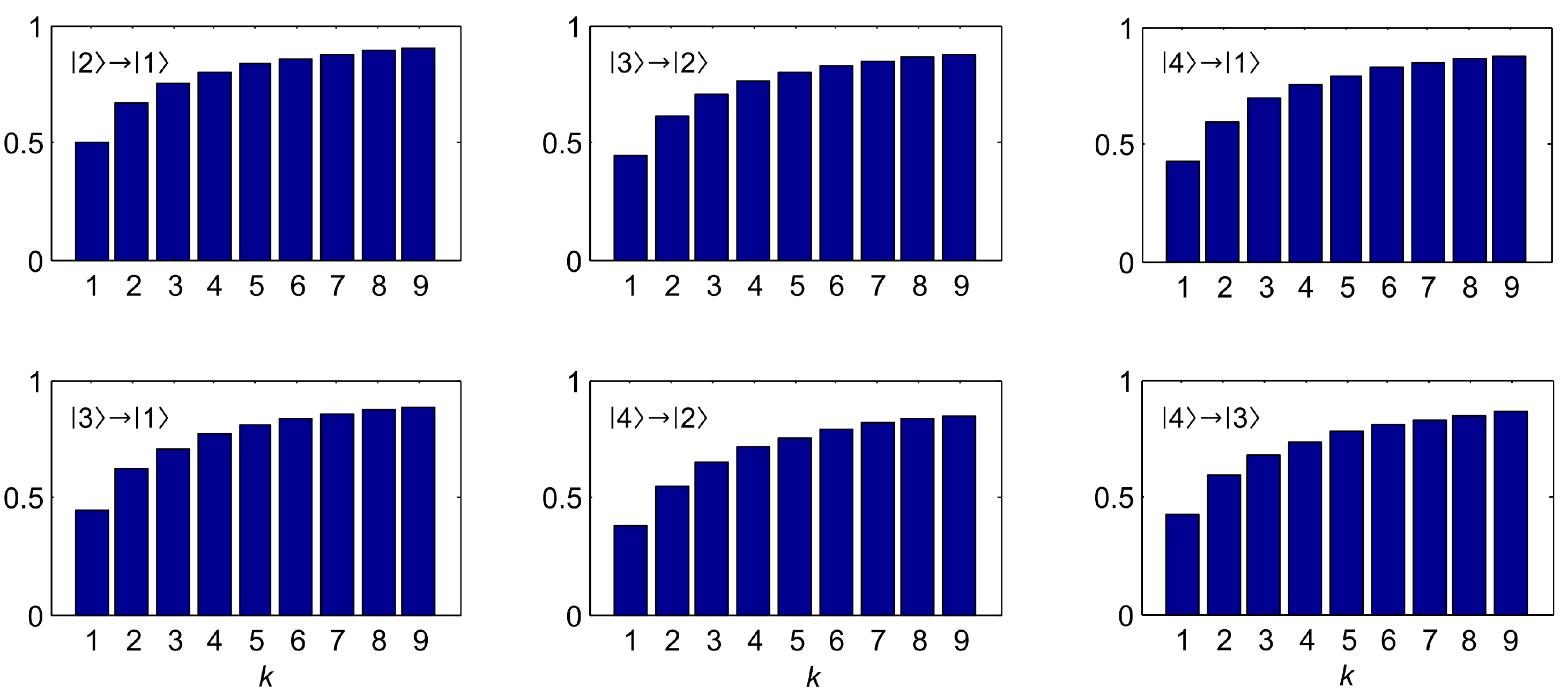}}
\caption{Optimal photon conversion probabilities $P_{\mathrm{max}}(m,n|k)$ are plotted for $6$ combinations of $m$ and $n$.}
\end{figure}

Imperfect single photon detectors with non-unit detection efficiency $\eta$ will lead to production of mixtures of various Fock states. The errors in conditional single 
photon subtraction due to imperfect detection can be reduced by using highly unbalanced beam splitter at the expense of reducing the overall success probability of the protocol. Since we here instead study the regime 
that maximizes the probability of state conversion, imperfect detection will unavoidably play a role. Let us illustrate this on the above considered example of $|2\rangle\rightarrow |1\rangle$ conversion. 
Remarkably, the normalized conditional
output state $\rho_{\mathrm{out}}$ is the same for both single-step conversion with $T_1=\frac{1}{2}$ and two-step conversion with $T_1=\frac{2}{3}$ and $T_2=\frac{1}{2}$ or $T_2=1$,
\begin{equation}
\rho_{\mathrm{out}}=\frac{1}{2-\eta}|1\rangle\langle 1 | + \frac{1-\eta}{2-\eta}|0\rangle\langle 0|.
\end{equation}
The overall probabilities of state preparation for the single- and two-step schemes exhibit similar dependence on $\eta$,
\begin{equation}
P(2,1|1,\eta)=\frac{1}{2}\eta(2-\eta), \qquad P(2,1|2,\eta)=\frac{2}{3}\eta(2-\eta),
\end{equation}
however with different prefactors. This shows that the advantage of feedforward-based scheme is preserved even for inefficient detection. Specifically, for a given output state quality we may achieve higher 
state preparation probability with feedforward.

Since the feedforward-controlled switchable beam splitter will in practice introduce additional optical losses, we now investigate a more refined model of the setup in Fig. 1(b) with a lossy channel 
with transmittance $\eta_O$ inserted in between the beam splitters $\mathrm{BS}_1$ and $\mathrm{BS}_2$, see inset in Fig. 3(b). We assume that $\eta_O$ is constant and does not depend on the setting of the transmittance of $\mathrm{BS}_2$, 
which is well justified  for setups based on interferometric schemes with feedforward-controlled electrooptic modulators. 
In order to compare the feedforward-based scheme with the elementary photon subtraction block in Fig. 1(a), 
we again consider the $|2\rangle\rightarrow |1\rangle$ Fock state conversion and we investigate the trade-off between the state conversion probability $P(2,1)$ 
and the single-photon fraction $p_1$ in the conditionally generated output 
state $\rho_{\mathrm{out}}=p_1|1\rangle\langle 1|+(1-p_1)|0\rangle \langle 0|$. For the elementary photon subtraction block in Fig.~1(a) we get a simple parametric dependence of $P(2,1)$ and $p_1$ 
on the transmittance $T$ of beam splitter BS,
\begin{equation}
P(2,1)=2\eta(1-T)[1-(1-T)\eta], \qquad p_1=\frac{T}{1-(1-T)\eta}.
\end{equation}
For $\eta\geq \frac{1}{2}$ the maximum achievable probability of state conversion is $\frac{1}{2}$, achieved for $T=1-\frac{1}{2\eta}$. At this point, $p_1=2-\frac{1}{\eta}$. 
For $\eta<\frac{1}{2}$ the maximum probability reads $2\eta(1-\eta)$, 
which is however approached in the undesirable limit $T\rightarrow 0$, when also $p_1\rightarrow 0$. On the other hand, in the limit $T\rightarrow 1$ also $p_1\rightarrow 1$ 
but at the cost of vanishing success probability, $P(2,1)\rightarrow 0$, a well-known limit of the standard photon subtraction scheme. The trade-off between P(2,1) and $p_1$ for the elementary photon subtraction block 
is plotted in Fig. 3 as blue solid line for two different $\eta$. The choice $\eta=60\%$ corresponds to detection of photons with (an array of) ordinary avalanche photodiodes while $\eta=0.85\%$ 
illustrates the performance for highly efficient detectors such as superconducting single photon detectors.

\begin{figure}[t!]
\centering\includegraphics[width=\linewidth]{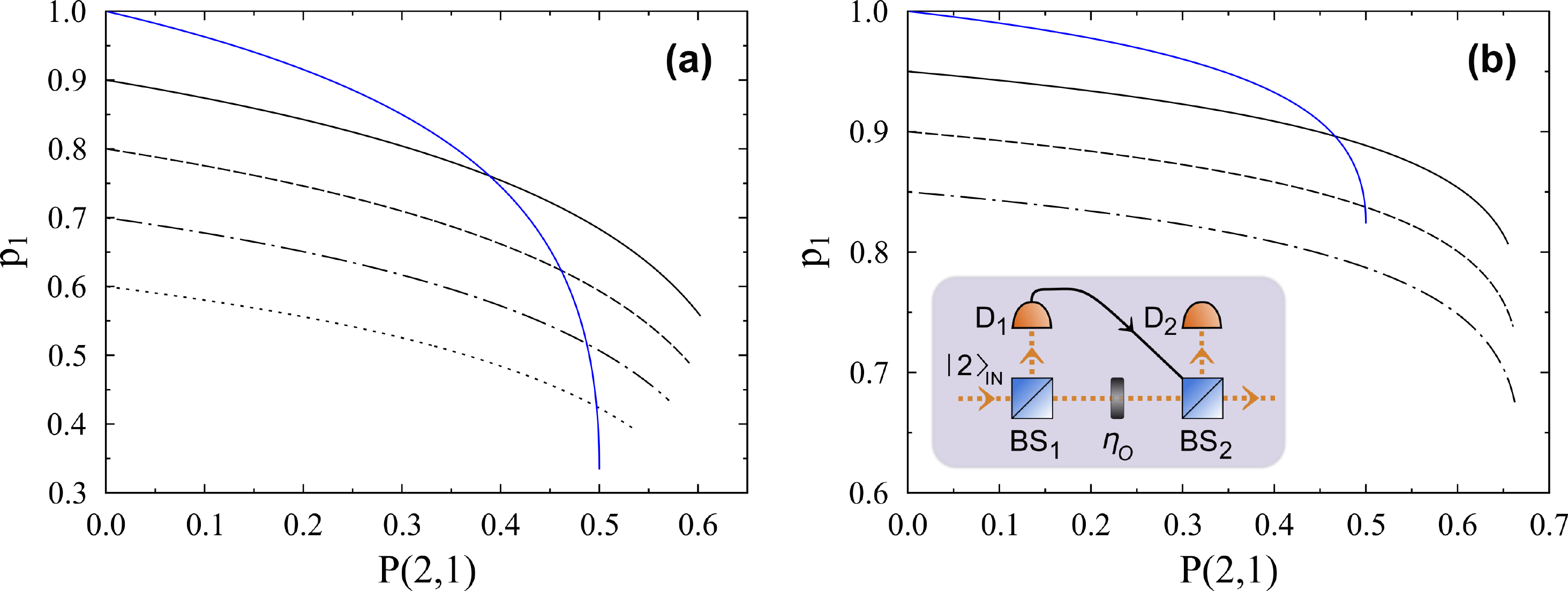}
\caption{Trade-off between probability of $2\rightarrow 1$ Fock state conversion $P(2,1)$ and the single-photon fraction $p_1$ in the output state is displayed for two different detection efficiencies $\eta=0.6$ (a) 
and $\eta=0.85$ (b).  The blue solid line represents result for the elementary photon subtraction block in Fig. 1(a). Black lines show results of the feedforward-based scheme with additional optical losses 
quantified by the effective transmittance $\eta_0$. The various lines in panel (a) correspond to  $\eta_O=0.9$ (solid line), $0.8$ (dashed line), $0.7$ (dot-dashed line) and $0.6$ (dotted line) while the lines on panel (b) 
are plotted for $\eta_O=0.95$ (solid line), $0.9$ (dashed line) and $0.85$ (dash-dotted line). The inset in panel (b) shows the optical model of the feedforward-based scheme with included optical losses of the switchable beam splitter $\mathrm{BS_2}$.}
\label{fig:losses}
\end{figure}

For the feedforward-based scheme with additional optical losses where the transmittance of $\mathrm{BS}_2$ is actively switched between $T_2$ and $1$, we obtain
\[
P(2,1)=2\eta\left\{ 1-T_1-\eta+\eta T_1[2-T_1-\eta_O(1-T_2)(1-T_1+\eta_OT_1(1-T_2))]+\eta_OT_1(1-T_2) \right\},
\]
\begin{equation}
p_1=\frac{\eta_O T_1[1-T_1+\eta_O T_1 T_2(1-T_2)]}{1-T_1-\eta+\eta T_1[2-T_1-\eta_O(1-T_2)(1-T_1+\eta_OT_1(1-T_2))]+\eta_OT_1(1-T_2)}.
\end{equation}
For any given $\eta$, $\eta_O$ and target conversion probability $P(2,1)$ we numerically optimize $T_1$ and $T_2$ to achieve maximum single-photon fraction $p_1$ in the output state. 
The results of numerical optimization are plotted in Fig. 3 for two different detection efficiencies $\eta$ and several different levels of added optical loss. The maximum achievable 
single-photon fraction is limited by the additional optical losses, $p_1 \leq \eta_O$. If $\eta_O(3\eta-1)\geq 1$, then the maximum conversion probability reads $P_{\mathrm{max}}(2,1)=\frac{2}{3}$, and is obtained for 
\begin{equation}
T_1=\frac{3\eta-1}{3\eta} ,\qquad T_2=1-\frac{1}{\eta_O(3\eta-1)}.
\end{equation}
The single photon fraction achieved at this point reads
\begin{equation}
p_1=2\eta_O-\frac{1+2\eta_O}{3\eta}.
\end{equation}
The graphs in Fig.~3 indicate that the feedforward-based scheme becomes advantageous provided that $\eta_O \gtrsim \eta$, 
i.e. the additional optical losses should be comparable to or smaller than the effective losses in single-photon detection. In our proof-of-principle fiber-based experiment at $810$ nm reported below, 
the losses imposed by electrooptic modulators and other optical components result in $\eta_O \lesssim 20\%$. However, an alternative approach using free-space electrooptic modulator based switches could reduce 
losses below $10\%$, yielding $\eta_O \gtrsim 90\%$,  making the scheme applicable in practice. 
 
\section{Experiment}
In this Section we report on a proof-of-principle experimental demonstration of feedforward-enhanced $|2\rangle \rightarrow |1\rangle$ Fock state conversion as sketched in Fig. 1(b).
The main aim of our experiment is to verify the feasibility of feedforward-controlled photon subtraction
and demonstrate the potential advantage of feedforward-based scheme in comparison to the elementary single-step photon subtraction block in Fig. 1(a). 
The experiment is therefore designed such as to emulate a perfect lossless setup for both schemes. 
We overcome the optical losses and finite detection efficiency by effectively balancing the losses in all channels and
measuring two-photon coincidence events that indicate either success or failure of the Fock state
conversion. We thus postselect only the cases when both input photons reach the single photon detectors and are detected.
In this approach, the overall losses are factored out and cancelled in the calculation of
the effective success probabilities of both schemes, that are determined as ratios of the measured
two-photon coincidence counts.
 Additionally, the projection onto the two-photon subspace enables us to emulate the two-photon source with a highly attenuated coherent state. 

\begin{figure}[t!]
\centering\includegraphics[width=\linewidth]{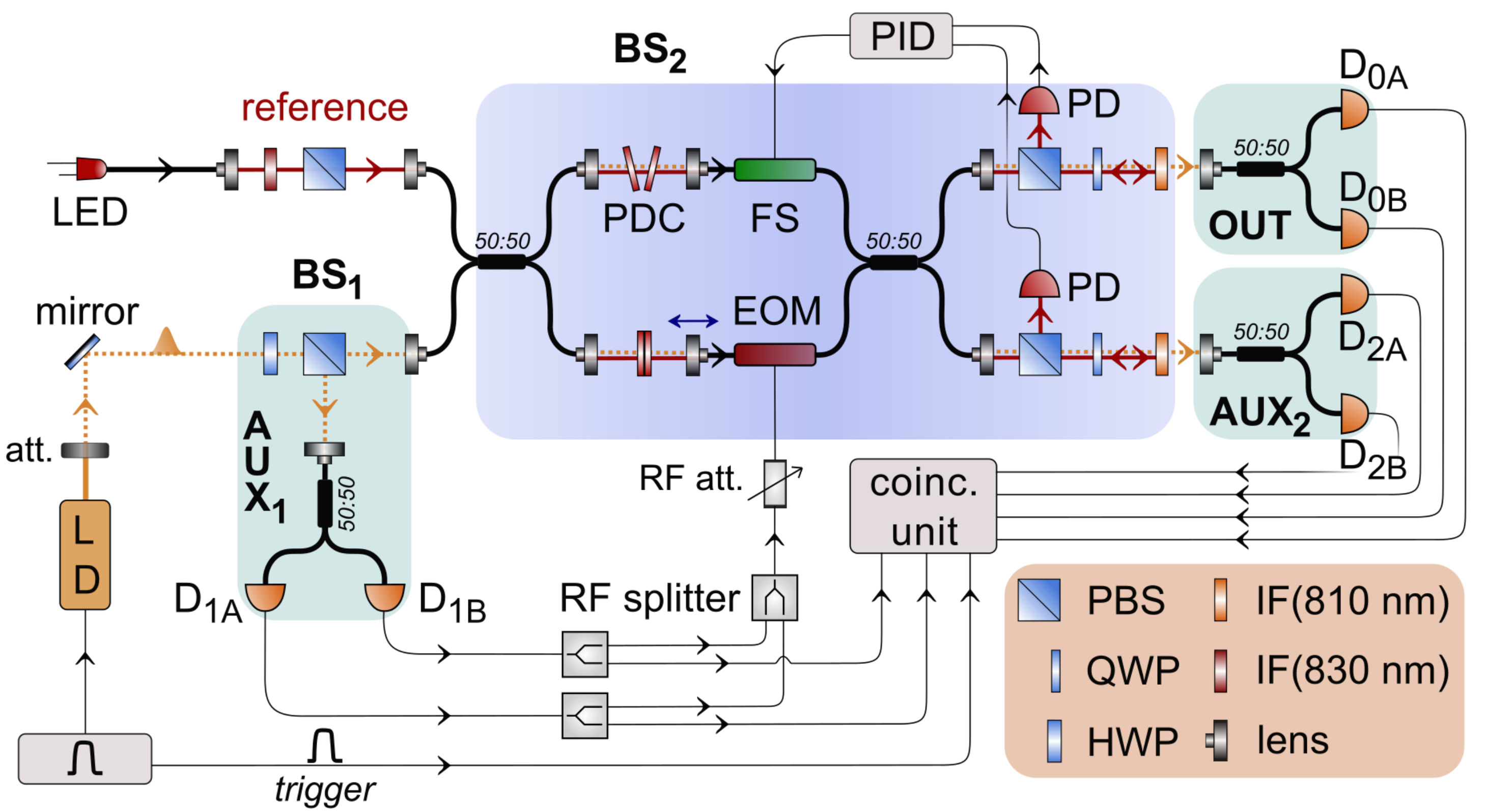}
\caption{Experimental setup for feedforward-enhanced $\ket{2}\rightarrow\ket{1}$ Fock state conversion. The main optical parts of the setup  include a variable beam splitter $\mathrm{BS}_\mathrm{1}$ and an electrooptically
switchable beam splitter $\mathrm{BS}_\mathrm{2}$ implemented as Mach-Zehnder interferometer. The feedforward control of $\mathrm{BS}_2$ is triggered by detection of a photon in the output port AUX$_1$. Legend: laser diode (LD), 
half-wave plate (HWP), polarization beam splitter (PBS), phase dispersion compensator (PDC), fiber stretcher (FS), electro-optic phase modulator (EOM), quarter-wave plate (QWP), photodiode (PD), 
single-photon detector (D), interference filter (IF).}
\label{fig:setup}
\end{figure}

Our experimental setup is depicted in Fig. \ref{fig:setup}. An attenuated laser diode periodically driven by 1~ns pulses at 2~MHz repetition rate produces signal photons at 810~nm. 
The signal passes through tunable beam splitter $\mathrm{BS}_\mathrm{1}$ realized as a sequence of a half-wave plate and a polarization beam splitter. 
Subsequently, the signal enters switchable beam splitter $\mathrm{BS}_\mathrm{2}$ implemented as a Mach-Zehnder interferometer (MZI) with a 10~GHz integrated electro-optic phase modulator PM-0K5-10-PFU-PFU-810-UL from EOSpace (EOM), 
enabling low-voltage and low-latency switching performance \cite{Svarc2019}. While the phase control is performed in an 8~m long MZI part formed of polarization-maintaining fibers, 
precise arm balancing, and dispersion compensation are done in 1~m long air gap. As a result, we reach the interference visibility 99.55\% enabling the switching with extinction greater than 400:1. 
The MZI has an overall transmission of $\sim20\%$. Due to the presence of environmentally induced phase fluctuations in MZI, an active phase-lock is implemented. It exploits auxiliary light 
at 830~nm acting as a phase reference. Particularly, we use a single-mode coupled luminescent diode with additional polarization and spectral filtering resulting in 3~nm bandwidth and power of 100~pW. 
The reference and the signal are merged and co-propagate through the MZI. Subsequently, at the outputs, the wavelengths are separated with a sequence of a polarizing beam splitter, a quarter-wave plate, 
and an interference filter acting together as an optical isolator. Phase fluctuations are monitored with ultra-sensitive photodiodes, evaluated with a custom-made analog proportional-integral-derivative (PID) controller, 
and compensated with a fiber stretcher performing at 1~kHz bandwidth and dynamic range of 35~$\mu$m. Our technique provides continuous tunability and sub-degree stability of the phase.

To provide the projection into two-photon subspace we need to discriminate at least between Fock states $\ket{2}$ and $\ket{1}$ at each output. 
We simplify generic photon number resolving detection by splitting the signal between two silicon avalanche photo-diodes with detection efficiencies around~$65\%$. 
However, compared to the ideal photon number resolving detection we can discriminate the Fock state $\ket{2}$ only with 50\% probability, which we take into account in data analysis. 
To provide balanced detection we optimize the setup to achieve equal click probabilities for 
neighbouring detectors $\mathrm{D}_{j\mathrm{A}}$ and $\mathrm{D}_{j\mathrm{B}}$.
Since loss in $\mathrm{AUX_1}$ port is much smaller than loss induced by $\mathrm{BS_2}$ we modify the splitting ratio of $\mathrm{BS_1}$ in order to compensate the imbalance. 
This approach is equivalent to imposing an additional artificial loss in the $\mathrm{AUX_1}$ port. It is then relevant to evaluate the effective splitting ratio of $\mathrm{BS}_1$ 
as the ratio of the signal detected at $\mathrm{AUX_2}$ and OUT ports to signal detected at $\mathrm{AUX_1}$ port.  Particularly, 
we determine the effective transmittance $T_\mathrm{eff}$ of $\mathrm{BS}_1$ as $(N_\textrm{0A}+N_\textrm{0B}+N_\textrm{2A}+N_\textrm{2B})/(N_\textrm{0A}+N_\textrm{0B}+N_\textrm{1A}+N_\textrm{1B}+N_\textrm{2A}+N_\textrm{2B})$ 
where $N_{i}$  denotes count rate at detector $\textrm{D}_{i}$. In our experiment, $T_\mathrm{eff}$ acts the same way as $T_1$ in the lossless case described by Eq. (\ref{eq:21conversion}).
To trigger the feedforward control of $\mathrm{BS}_\mathrm{2}$, 
electronic pulses generated by $\mathrm{D}_\mathrm{1A}$ and $\mathrm{D}_\mathrm{1B}$ are utilized. The pulses are merged, set to $\pi/2$ modulation voltage and fed into EOM. 
To reach the same shape and timing of the pulses, discriminators and delay lines are used (not shown in the scheme). 

\begin{figure}[t!]
\centering\includegraphics[width=\linewidth]{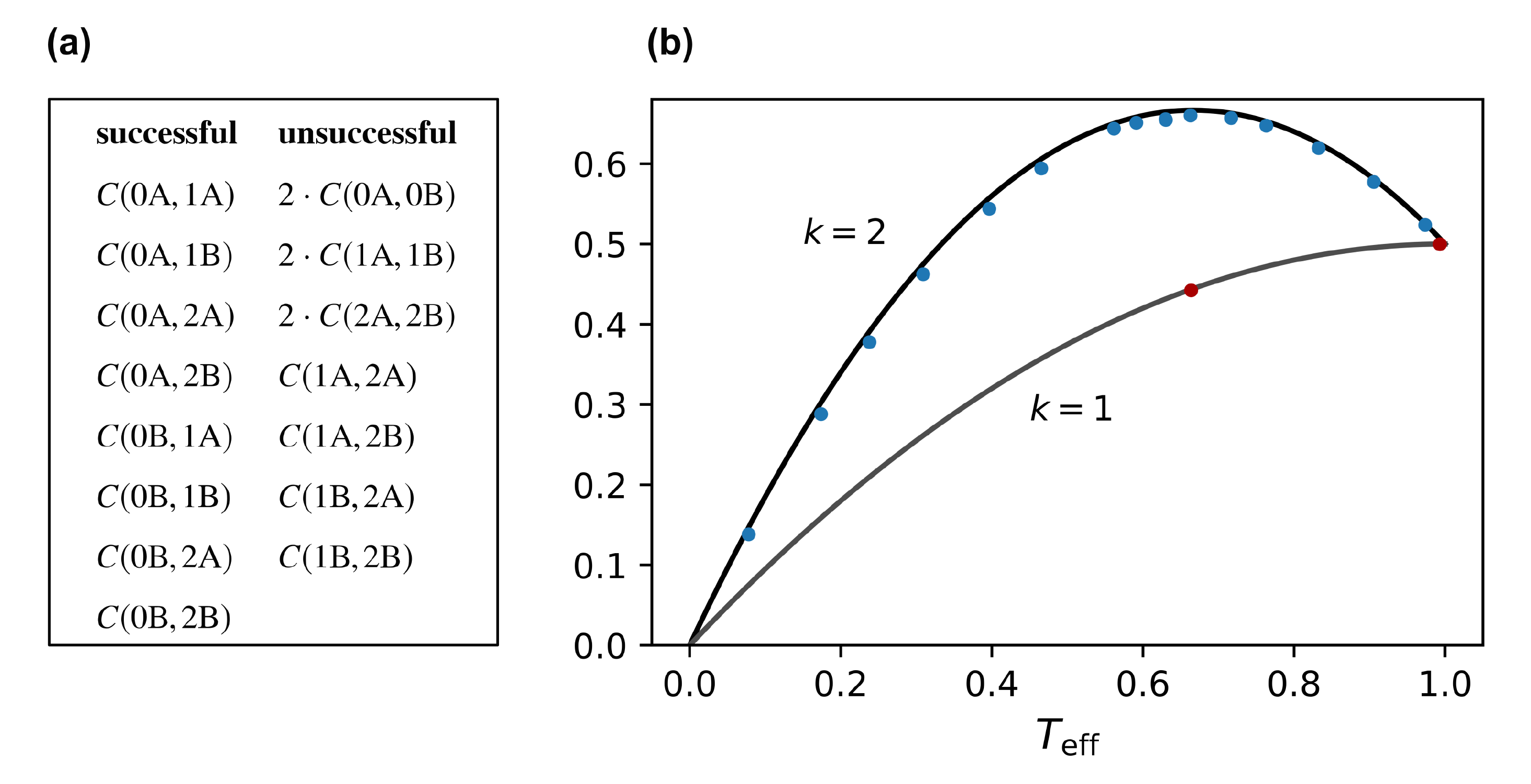}
\caption{(a) Table of coincidence tagging, where $C(i, j)$ denotes a coincidence event between detector $\mathrm{D}_i$ and $\mathrm{D}_j$. 
For neighbouring detectors an event is counted twice, because if two photons reach the same output port AUX$_1$, AUX$_2$ or OUT, they trigger the coincidence event only with 50\% probability. 
(b) Experimental results of conversion probability $P_\mathrm{exp}(2,1|k)$ for $k=1,2$ depending on effective spitting ratio $T_\textrm{eff}$. 
Blue dots representing data of feedforward-enhanced conversion $k=2$ are plotted against the single beam splitter conversion $k=1$ shown as red dots. 
Black and grey lines represent upper bounds for the ideal conversion. Error bars are smaller than the point size.}
\label{fig:vysledek} 
\end{figure}

For data collection and processing, a custom made 16-channel coincidence 
unit is utilized \cite{Hlousek2019}. To avoid random detection events, the measurement is triggered by the laser diode driving pulse. 
We detect all of the possible 15 combinations of two-coincidences as listed in Fig. \ref{fig:vysledek}(a). 
A coincidence event is tagged as \textit{successful} if one photon of the pair is detected at the output signal port OUT, while the other photon is heralded at $\mathrm{AUX_1}$ or $\mathrm{AUX_2}$ port. 
All other coincidence events are tagged as \textit{unsuccessful}. Events discriminated only with 50\% probability are counted with double rate. 
The effective success probability of conversion is then determined as the ratio of successful coincidence counts to all coincidence counts.
The input coherent state contains a small amount of higher photon-number states that may be falsely indicated as two-coincidences and influence the results. 
Evaluating higher-order coincidences we estimate that spurious two-coincidences form $\sim$1\% of the signal causing a relative error of 0.4\% in the worst case. 
Further reduction of the error is achievable by additional attenuation of the signal source, however, at the expense of a decreased rate of the two-photon state.

Results of feedforward-enhanced $\ket{2}\rightarrow\ket{1}$ conversion are shown in Fig. \ref{fig:vysledek}(b). For comparison, we include data of single beam splitter conversion achieved by deactivation of the feedforward.
According to Eq. (\ref{eq:21conversion}), the best possible performance $P_{\mathrm{max}}(2,1|2)=66.7\%$ is predicted for $T_\textrm{1}=66.7\%$. Experimentally we reach a very close value of 
$P_{\mathrm{exp}}(2,1|2)=(66.0\pm0.1)\%$ for $T_\textrm{eff}=(66.30\pm0.05)\%$. Error caused by spurious coincidences is estimated as $0.2\%$. 

Our results show that the proposed protocol can work experimentally with nearly ideal performance. Extension to arbitrary $\ket{m}\rightarrow\ket{n}$ conversion is possible, 
provided additional detection multiplexing would be used. To improve the probability of success, an extension of the experimental setup up to $k$ beam splitters would be necessary. 
A resource-efficient approach would be reusing single beam splitter for $k$ times in a loop \cite{Calsamiglia2001,Marek2018}. 
Although our fiber-based experimental setup is not suitable for cascading due to a high amount of loss, an alternative approach using free-space electrooptic modulator based switches would be convenient. 
The loss can be reduced below 10\% per cycle, making it practical even without the need of a coincidence basis. 
The limitation of this approach is the higher latency of the feedforward estimated to tens of nanoseconds ultimately \cite{Zeilinger2007,Zeilinger2011,Pan2017}.

\section{Conclusion}
In summary, we have proposed and experimentally demonstrated a feedforward-enhanced scheme for optical Fock state conversion via photon subtraction. In our approach, the transmittances of  tapping beam splitters are controlled by all preceding measurement outcomes to maximize the success probability of photon subtraction for a given setup complexity, i.e. a given maximum number of elementary photon subtraction blocks. Our results for $|2\rangle \rightarrow |1\rangle$ conversion are directly applicable to single photon subtraction from a single-mode weakly squeezed vacuum state that can be approximated as $|0\rangle+\epsilon|2\rangle$, since the dominant vacuum term does not contribute to the subtraction. Our findings clearly demonstrate the usefulness and advantages of the presented feedforward-based method, which can be utilized for an arbitrary input state. Note however, that for input superpositions of Fock states, the output state would depend on the overall transmittance of the beam splitters hence a mixed state would be generated even with perfect detection. In such case one could investigate the trade-off between state preparation quality, as quantified e.g. by state fidelity, and success probability of the protocol and choose the most suitable operating point.

\section*{Funding}
Czech Science Foundation (project 19-19189S);
Palack\'y University (project IGA-PrF-2019-010).
\section*{Acknowledgments}
We thank Michal Dudka for the development and implementation of custom electronics.
\section*{Disclosures}
The authors declare that there are no conflicts of interest related to this article.

\end{document}